# A morphotropic phase boundary system based on polarization rotation and polarization extension

Dragan Damjanovic[a]

*Ceramics Laboratory, Swiss Federal Institute of Technology in Lausanne - EPFL,*

*1015 Lausanne, Switzerland*

*Abstract* – Many ferroelectric solid solutions exhibit enhanced electro-mechanical properties at the morphotropic boundary separating two phases with different orientations of polarization. The mechanism of properties enhancement is associated with easy paths for polarization rotation in anisotropically flattened free energy profile. Another mechanism of properties enhancement related to free energy flattening is polarization extension. It is best known at temperature-driven ferroelectric-paraelectric phase transitions and may lead to exceedingly large properties. Its disadvantage is temperature instability of the enhancement. In this paper a temperature-composition phase diagram is proposed that exhibits compositionally driven-phase transitions with easy paths for both polarization rotation and polarization extension.

PACS: 77.80.B-, 77.80.bg, 77.84. Cg,

[a] e-mail: dragan.damjanovic@epfl.ch

,

The properties of materials that are defined by polarization change, such as dielectric permittivity and piezoelectric coefficients, may be enhanced in phase transition regions where there is a significant polarization variation. The examples are divergence of the dielectric and piezoelectric properties near points of temperature[1]- (see Fig. 1), stress-[2], electric field-[3,4] and compositionally-[5,6] driven structural phase transitions. The compositionally-induced structural change, the so-called morphotropic phase transition,[6,7] is of a great practical interest because the variable that drives the transition (i.e., composition) is inherent to the material and the transition point is maintained at operating conditions without an external influence (e.g., temperature, electric field or stress). The morphotropic phase transition is the origin of large piezoelectric properties in $Pb(Zr_{1-x}Ti_x)O_3$ solid solution (PZT), the most extensively used piezoelectric material. The enhancement of the properties in PZT occurs in the region of the composition-temperature phase diagram where crystal structure changes from tetragonal (T) to rhombohedral (R) via an intermediary monoclinic (M) phase as the Zr/Ti ratio becomes greater than ~52/48 (see Fig. 2).[8] This region is known as the morphotropic phase boundary (MPB).[6,7] Search for new piezoelectrics (for example, lead-free) is naturally focused on materials exhibiting a PZT-like MPB. In this paper the general concept of properties enhancement in the phase transition regions is used to propose a modified type of the phase diagram with compositionally-driven phase transitions.

*Ab-initio* and phenomenological calculations have shown that the intrinsic mechanism of the properties enhancement in phase transition regions is flattening of a free energy profile.[2,3,9-13] This mechanism appears to be common to most phase transitions, whether compositionally-, stress-, electric field- or temperature- driven. The anisotropy of the free energy profile determines easy paths for polarization change

,

and corresponding properties enhancement.[11] Consider first a familiar example of the temperature-driven phase transitions in the T phase of barium titanate, $BaTiO_3$ (Fig. 1). Tetragonal barium titanate exhibits spontaneous polarization $P^T = (0,0,P_3^T)_C$ oriented along $[001]_C$ pseudocubic direction. For simplicity, here and in the rest of the text only one of several equivalent crystallographic directions is considered. At 393 K (the Curie temperature) $BaTiO_3$ transforms into a cubic (C) phase with zero polarization, $P^C = (0,0,0)_C$, while at 278 K it transforms into an orthorhombic (O) phase with polarization $P^O = (P_1^O,0,P_1^O)_C$ oriented along $[101]_C$ pseudocubic direction. A free energy flattening near the T-O phase transition temperature favors easy polarization rotation in the $(010)_C$ plane, from $(0,0,P_3^T)_C$ to $(P_1^O,0,P_1^O)_C$. The material coefficients related to such polarization change are enhanced as the transition temperature is approached (in this case transverse susceptibility $\chi_{11}$ and shear piezoelectric coefficient $d_{15}$). The transition from the T to C phase favors one-dimensional change of polarization from $(0,0,P_3^T)_C$ to $(0,0,0)_C$, i.e., collinear polarization contraction/extension along $[001]_C$ direction.[10,14] The enhanced material coefficients in this case are the longitudinal susceptibility $\chi_{33}$ and piezoelectric coefficient $d_{33}$ (Fig. 1). Note that $\chi_{11}$ also increases near the T-C phase transition temperature because the free energy profile becomes globally flatter, but $\chi_{11}$ diverges less than $\chi_{33}$. Hereafter, discussion is always limited to the dominant effect. Free energy profiles for tetragonal $BaTiO_3$ can be found in Refs. 9,10. For simplicity term "extension" is used for both polarization contraction and extension.[14]

Analogous situation takes place in PZT in the vicinity of the MPB and Curie point, except that the variable that drives the transition at MPB is composition and not temperature. Ambient temperature behavior of PZT, at 300 K, is first considered (see

,

Fig. 2). As MPB is approached from the rhombohedral side of the phase diagram, a free energy profile flattens favoring polarization rotation from $P^R = (P_1^R, P_1^R, P_1^R)_C$ [along the pseudocubic $(111)_C$ direction] toward $P^T = (0,0,P_3^T)_C$ [the pseudocubic $(001)_C$ direction] via monoclinic mirror plane $(\bar{1}10)_C$; the easy R-T polarization path is illustrated in Fig. 2 for rhombohedral PZT 60/40 composition.[11] Similar effect also takes place when MPB is approached from the tetragonal side, although the easy polarization path may be different than on the rhombohedral side.[11] Propensity of PZT compositions in the MPB region to polarization rotation has been now firmly established as the main intrinsic mechanism for the enhancement of the electro-mechanical properties.

Considering again the rhombohedral PZT 60/40 composition, the discussion is now moved to 540 K. Note that at this temperature, PZT 60/40 is close to the R-C transition (the Curie point) and R-T transition (MPB) and both are close to the triple point T-R-C where the three phases meet. As the temperature is increased the composition PZT 60/40 gets closer to the MPB because of the slight curving of the boundary, Fig. 2. The flattening of the energy profile favoring polarization rotation thus becomes even more pronounced. However, another feature of the energy landscape develops approaching the Curie point: as the C phase is approached on heating, the energy profiles between R and C phases on one side and T and C phases on the other side become flatter, as shown in Fig 2. This is analogous to the behavior of BaTiO$_3$ near its Curie point[10] except that properties of PZT near the T-R-C point are enhanced for two reasons. One is the MPB effect, which favors polarization rotation and enhancement of the transverse susceptibility and shear piezoelectric coefficients. The other is proximity of the Curie point, which favors enhancement of the longitudinal piezoelectric coefficient and longitudinal susceptibility. It should be noted that a free energy flattening along one direction in general affects paths along other directions. The author suggests that this two-dimensional flattening of the energy profile (polarization rotation due to MPB and polarization extension due to proximity of the T-C and R-C Curie points) contributes to the exceptionally large piezoelectric properties reported recently by Liu and Ren[15] in (Ba,Ca)(Zr,Ti)$O_3$ ceramics.

In the search for new materials with properties comparable to those of PZT it is obvious to concentrate on systems exhibiting a PZT-like MPB. The most studied solid solutions (e.g., BiFe$O_3$-PbTi$O_3$,[16] BiSc$O_3$-PbTi$O_3$,[17] Pb(Zn$_{1/3}$Nb$_{2/3}$)$O_3$-PbTi$O_3$[18]) are based on the same polarization rotation mechanism as in PZT. The possibility of properties enhancement through polarization extension is largely ignored. One reason for this is that in most ferroelectrics the easy polarization extension paths in a free energy profile are temperature-driven and are thus impractical. The second reason is that the polarization rotation is overwhelmingly accepted as the "only" mechanism for substantial enhancement of the piezoelectric properties. This opinion is held despite the fact that the polarization extension processes may lead to exceptionally large piezoelectric properties. In fact, the largest piezoelectric coefficient ever reported ($d_{16} \sim 20000$ pm/V, measured in $KH_2PO_4$[19]) is related to polarization extension, not polarization rotation.[20]

One way to benefit from the polarization extension mechanism without disadvantages of the temperature-driven instabilities would be in a solid solution that exhibits morphotropic phase boundary that separates a polar and a nonpolar phase. Such an MPB would be analogous to the T-C transition in BaTi$O_3$ or R-C and T-C transitions in PZT, but would be driven by composition, not temperature. Exactly such MPB has been recently discovered in nonferoelectric solid solution between polar AlN and nonpiezoelectric ScN.[21,22] The piezoelectric $d_{33}$ coefficient of Al$_{1-x}$Sc$_x$N exhibits an impressive 400-500% increase as the MPB is approached within the AlN –rich side of the phase diagram. Importantly, it has been shown that the origin of piezoelectric enhancement in $Al_{1-x}Sc_xN$ is the same as in ferroelectric systems, i.e. flattening of a free energy profile. In this case the flattening favors polarization extension path, confirming the general nature of the mechanism.[21] While this manuscript was in preparation Datta *et al.*[23] have reported a compositionally-driven transition from a paraelectric into a ferroelectric phase in $BiInO_3$-$BaTiO_3$ solid solution, but properties have not been measured. Other systems that may be worth examining for presence of MPB with compositionally-driven polarization extension mechanism are solid solutions based on $BiFeO_3$.[24,25]

After establishing that compositionally-driven phase transitions with polarization extension easy paths are reality it needs to be mentioned that properties at such MPB would benefit only from polarization extension but not, in general, from the polarization rotation. This is clearly not sufficient. In polycrystalline materials, for example, only some grains would be favorably oriented with respect to the field to take a full advantage of the polarization extension and the effect on properties enhancement is not expected to be high. One should, therefore, look for a solid solution exhibiting MPBs with easy paths for *both* polarization rotation and polarization extension. The phase diagram shown in Figure 3 is proposed as a general representative of such a solid solution: the phase diagram exhibits a narrow polar phase (2-3 mol% wide) with T or M symmetry surrounded on one side by a polar R or O phase and on the other side by a nonpolar phase. The transition temperature into the high temperature paraelectric phase should be high. Because of the narrowness of the intermediate T/M phase, properties of such a system would benefit from both polarization rotation (T/M↔R/O) and polarization extension (T/M↔nonpolar phase) mechanisms without disadvantages of

,

the thermal instability usually associated with temperature-driven polarization extension mechanism. Clearly, T/M and R/O regions may switch positions, but the nonpolar region should not be in the middle.

Two points about the proposed phase diagram may be immediately emphasized. Large properties in MPB systems (e.g., PZT) are only partly due to intrinsic structural instabilities discussed above. The other, equally strong if not stronger contribution to the properties has origin in domain wall motion. Domain wall contributions depend on crystallographic phases[26] on each side of the MPB and it will be interesting to see whether the phase diagram shown in Fig. 3 permits significant domain wall contribution to the electro-mechanical properties. Another point is that the nonpolar phase should probably not be antiferroelectric, such as $PbZrO_3$. The reason for this is that polarization extension and rotation mechanisms operate on unit cell scale. Even though macroscopic polarization of an antiferroelectric phase is zero, each unit cell exhibits nonzero polarization so that the polarization extension mechanism may not fully develop.

In summary, a phase diagram is proposed with a morphotropic phase boundary region benefiting from both compositionally-driven polarization rotation and polarization extension mechanisms.

The author acknowledges Swiss National Research Program on Smart Materials for financial support (PNR 62, Contract No. 406240-126091).

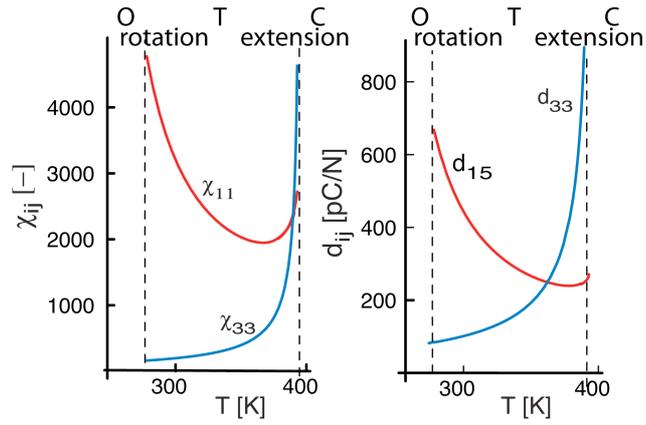

Figure 1, Damjanovic APL

Figure 1. (color online) The dielectric susceptibility χ and piezoelectric coefficients *d* in the tetragonal phase of BaTiO$_3$ calculated as a function of temperature using Landau-Ginzburg-Devonshire theory. For details see Ref. 1. The vertical lines designate transition temperatures into orthorhombic and cubic phases. BaTiO$_3$ exhibits strong propensity toward polarization rotation near the T-O and toward polarization elongation close to the T-C phase transition temperature.

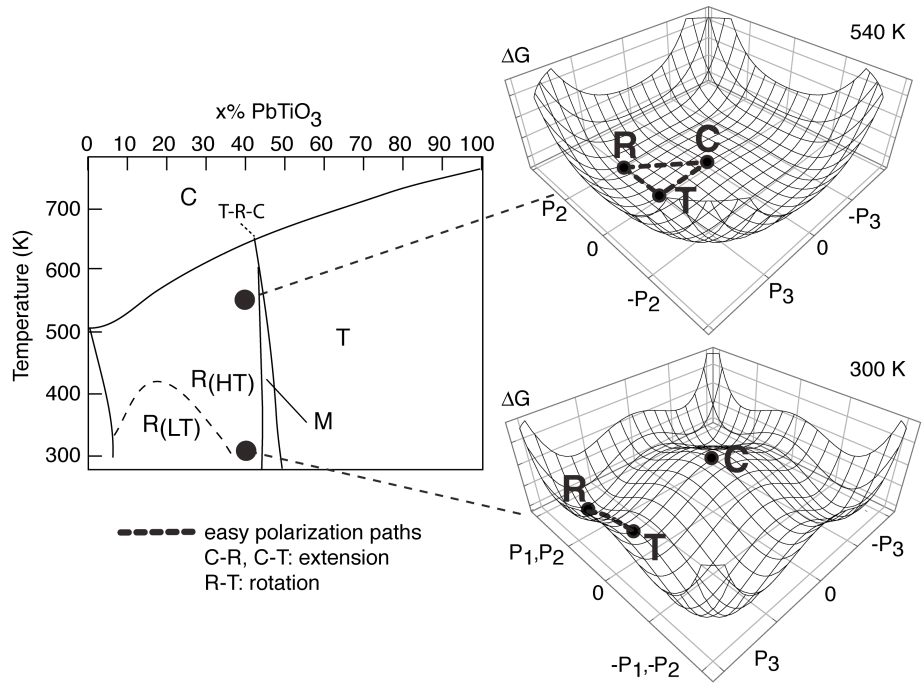

Figure 2, Damjanovic APL

Figure 2. The composition-temperature phase diagram of PZT (after Ref. 6,8) and Gibbs free energy profiles for PZT with Zr/Ti ratio 60/40 calculated at 300 K and 540 K using Landau-Ginsburg-Devonshire theory (Refs. 5,11). The black dots in the phase diagram mark the temperatures at which the free energy profile was calculated. The black dots in the energy profiles mark equilibrium rhombohedral phase with polarization $(P_1^R, P_1^R, P_1^R)_C$ and tetragonal and cubic phases with polarizations $(0,0,P_1^R)_C$ and $(0,0,0)_C$, respectively. The thick dashed lines indicate easy polarization paths at the two temperatures. T-C and R-C paths indicate polarization extension and R-T path indicates polarization rotation path.

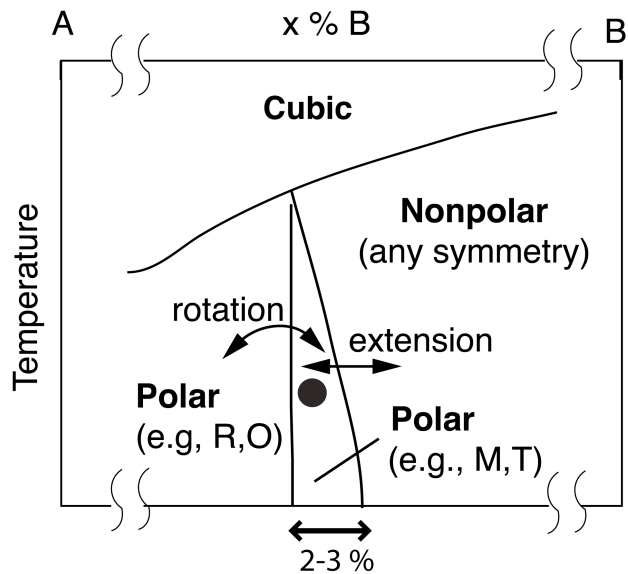

Figure 3, Damjanovic APL

Figure 3. A hypothetical phase diagram exhibiting an MPB region where both polarization rotation and polarization extension mechanisms may be strong. The mutual sequence of R/O and T/M phases may not be important, however the nonpolar phase should not be in the middle of the diagram. The full dot represents composition that benefits from both polarization rotation and polarization extension at temperatures well below Curie or triple point.